\documentclass[twocolumn,showpacs,preprintnumbers]{revtex4} 
\usepackage{amssymb}
\usepackage[dvips]{graphicx}

\begin{document}

\title{Comment on `Intrinsic tunnelling spectroscopy of Bi$_2$Sr$_2$CaCu$_2$O$_{8+\delta}$: The junction-size dependence of self-heating'[Phys.Rev.B 73, 224501 (2006)]}

\author{V.N. Zavaritsky$^{1,2}$}
\address
{$^{1}$Department of Physics, Loughborough University, Loughborough, United Kingdom, 
$^{2}$Kapitza Physics Institute \& General Physics Institute, Moscow, Russia\\
}

\begin{abstract}
The recent PRB 73, 224501 (2006) henceforth referred as Ref.\cite{0} asserts that self-heating decreases with sample area reduction and claims to identify the intrinsic cause of ITS in submicrometre `mesa'. I will show that this assertion lacks substantiation. I will further demonstrate that one and the same $R(T)$ and the parameter-free Newton's Law of Cooling describe quantitatively a rich variety of ITS behaviours taken by Ref.\cite{0} above and below $T_c$ at bath temperatures spanned over 150K. Thus this finding presents strong evidence in favour of heating as the cause of  the `intrinsic tunnelling spectra' (ITS) promoted by Ref.\cite{0}.



\end{abstract}

\maketitle

Heating is probably the most common problem in low temperature research and a harsh limiting factor for the study of current-voltage characteristics (IVC). Its signatures are particularly well studied for the case of metals and conventional superconductors; notably, heating is known to cause IVC nonlinearities and transform a single-valued IVC into a multi-valued characteristic with regularly spaced branches (see Ref.\cite{2} for review). Unlike conventional superconductors, high temperature superconductors (HTSC) reveal exceptionally poor thermal and electrical conductivities which makes them particularly prone to local heating. The heating-induced IVC nonlinearities in such circumstances were found to resemble true tunnelling characteristics so closely that one may easily fall into the trap of identifying HTSC material as an SIS Josephson junction (see Ref.\cite{1} for review). 
The IVC by the authors of \cite{0} reproduced in Fig.1(a) are remarkably similar to extrinsic ones; however they are argued to be virtually free of heating. Below I will address the consistency of the basic assumptions, the experimental data and the conclusions by the authors of \cite{0}.




Assuming that self-heating in samples of different area is proportional to heat $W$=$IV$ only, the authors of Ref.\cite{0} claim that self-heating can be significantly reduced by means of sample area reduction. However, this assumption is incorrect as heat $W$, dissipated in a sample, escapes through its surface area $A$, so the temperature rise depends on the heat load $P=W/A$. Hence this claim lacks grounds and, moreover, is at odds with experimental data which point to the area independence in heating effects, see Ref.\cite{1} for review. Additional evidence in support of this conclusion may be seen in Ref.\cite{4}, which finds that practically the same heat loads $P\sim10kW/cm^2$ build the ITS gap in the mesas of vastly different area $1<A<30\mu m^2$ made of the same Bi2212 crystal. \footnote{It is worth noting that heating does not depend on A under otherwise identical experimental conditions. However this is not always the case in real `mesas' where the heat escapes into the bath primarily through the topmost metal electrode. As convective heat transfer depends on electrode geometry and area, the heat transfer coefficient might not necessarily remain exactly the same.} 
As far as temperature rise is concerned, the systematic experimental studies summarised in Ref.\cite{1} suggest that the mean temperature, $T$,  of the self heated sample is appropriately described by Newton's Law of Cooling,

\begin{equation}
T=T_B+P/h, 
\end{equation}
where $T_B$ is the temperature of the coolant medium (liquid or gas) and $h$ is the heat transfer coefficient, which depends neither on $A$ nor $T$. Furthermore, \cite{1} shows that {\it in layered HTSC heating-induced IVC nonlinearities exceed the intrinsic ones so radically that the latter might be safely ignored even at quite modest overheatings}. This finding, also supported by the experimental data in \cite{0}, is of particular importance to the present discussion. 

Although the assumptions by the authors of Ref.\cite{0} are not beyond dispute, it is worth considering their data, obtained in a state-of-art experiment. First I will verify the experimental consistency of \cite{0} using the IVC origins least affected by heating and thus allow comparison with R(T) data measured independently. As shown elsewhere, \cite{1}, in the absence of heating the initial slope of IVC taken at $T_B$ corresponds to normal state resistance, $R_N(T_B)$, which merges $R(T)$ if $T_B>T_c$. Indeed, close correlation between the thus determined $R_N(T_B)$ and  $R(T)$ measured by \cite{0} above $T_c$ is clear from Fig.1(b). This correlation provides strong evidence in support of the experimental consistency of the data by \cite{0}. Furthermore, in qualitative agreement with the direct measurements by \cite{asa,3}, $R_N(T_B)$ in Fig.1(b) continues its growth and retains its upward curvature when the temperature is lowered through and below $T_c$, as shown by the thin line in Fig.1(b). However, as  seen in Fig.1(b) deviations from this behaviour rapidly develop when $T_B$ lowers below 60K. As shown by Yasuda et al,\cite{yasuda} heating is a likely cause of such deviations, so there is a possibility that the lowest two $T_B$ were seriously underestimated (see the arrow in Fig.1(b)). Thus our analysis demonstrates the reasonable consistency of the experimental data by \cite{0} and suggests that in at least five of seven IVC by \cite{0} the quoted $T_B$ might be used as a reliable starting temperature in the analysis of heating issues.   

\begin{figure}
\begin{center}
\includegraphics[angle=-0,width=0.47\textwidth]{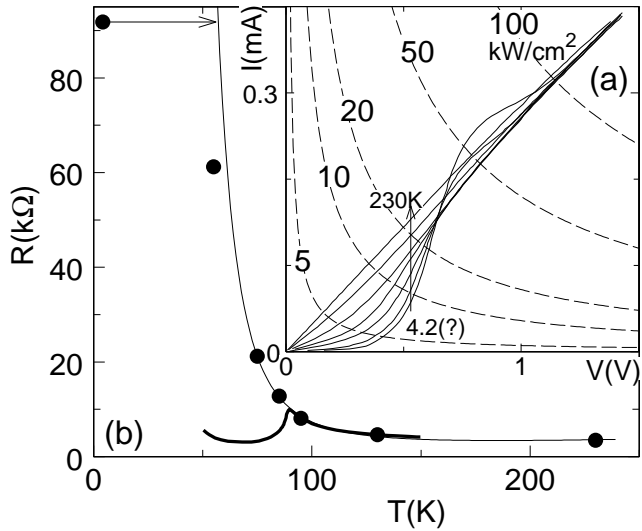}
\vskip -0.1mm
\caption{{\bf (a)}:  Solid lines reproduce nonlinear IVC, reported by Ref.\cite{0} for different $T_B$=4.2,55,75,85,95,130,230K above and below $T_c=89K$; broken lines show the levels of constant heat load $P$=$IV/A$=5,10,20,50,100$kW/cm^2$; $A$=$0.36\mu m^2$. {\bf (b)} compares measured R(T) shown by the thick solid line and $R(P$$\rightarrow$$0)$ vs $T_B$ estimated from the initial slopes of the IVC in Fig.1a.
}
\end{center}
\end{figure}

Thus, the experimental data by the authors of \cite{0} are sufficiently consistent and make it possible to address the origin of IVC nonlinearities using the parameter-free description by Ref.\cite{1}. These IVCs, which are central to the discussion in \cite{0}, are reproduced in Fig.1(a) together with the levels of constant heat load. As could be easily seen from this figure, in remarkable similarity with the basic ITS studies, nearly 10$kW/cm^2$ is required to build the characteristic IVC features attributed to the superconducting gap (ITS gap) by \cite{0}. Such loads exceed the critical ones by several orders, hence suggesting strong heating. However, additional analysis is required to quantify the temperature rise and, most importantly, to verify the extent to which the IVC nonlinearities might reflect the plausible intrinsic ones.

To discriminate between intrinsic and extrinsic contributions, let us compare the measured data with those calculated on the basis of the assumption that heating-induced IVC nonlinearities exceed the intrinsic ones so radically that the latter might be safely ignored. The IVC in such circumstances is primarily determined by $R_N(T)$, while the self heating is appropriately described by Eq.(1), see \cite{1} for details. To make such quantitative analysis feasible, the heat transfer coefficient, $h$ should be determined for the experimental conditions of \cite{0}. As shown in \cite{1} (and reaffirmed by independent measurements by \cite{7}), $h$ could be obtained  from a single self-heated IVC provided that the sample's R(T) is known. As will be shown in the next paragraph, the good data presentation by the authors of \cite{0} makes it possible to fulfil this task with even higher reliability. 

\begin{figure}
\begin{center}
\includegraphics[angle=-0,width=0.47\textwidth]{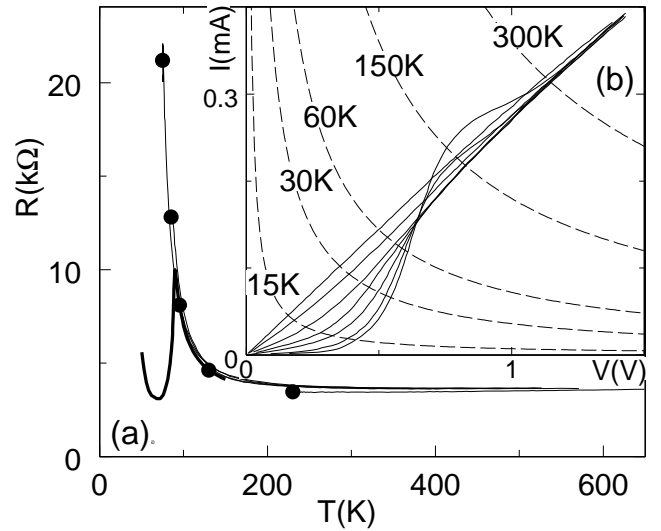}
\vskip -0.1mm
\caption{{\bf (a)}: Compares measured $R(T)$ shown by the thick solid line with those calculated with Eq.(1) from five of seven nonlinear IVCs from Fig.1(a) using  the same heat transfer coefficient  $h=350Wcm^{-2}K^{-1}$ for the data taken at $T_B$ spanned over 150K; solid dots represent corresponding $R(P$$\rightarrow$$0)$ vs $T_B=75,85,95,130,230K$.  {\bf (b)}: Solid lines reproduce the IVC from Fig.1(a); broken lines show the levels of constant overheating $\Delta T$=$T-T_B$=15,30,60,150,300$K$; $A$=$0.36\mu m^2$.   
}
\end{center}
\end{figure}

In terms of heating issues, it is appropriate to consider $R=V/I$ as a function of heat load, $P = V I/A$, rather than IVC only (see above and also Ref.\cite{1}). Provided that $T_B$ is correct, the thereby determined $R(P)$ could be converted with Eq.(1) into $R(T)$ thus giving a reliable estimate of $h$. The set of drastically different IVC of the same sample at various $T_B$ reported by \cite{0} provides a harsh consistency check for our approach as one and the same $h$ should convert various IVC into a single $R(T)$ thus leaving absolutely no space for manoeuvre. 
However, as seen from Fig.2(a), the parameter-free Eq.(1) collapses all IVCs obtained at $T_B$ spanned over 150K into a single curve which reproduces quantitatively the $R(T)$ of the same 'mesa' and allows estimate of the heat transfer coefficient $h$=(300-350)$Wcm^{-2}K^{-1}$. 
Thus, Fig.2(a) confirms the heating origin of the IVC non-linearity and suggests that the IVC by Ref.\cite{0} will be almost linear above and below $T_c=89K$ if the heating artefacts are removed. 

In addition it is worth considering the remaining two IVCs measured in Ref.\cite{0} in this sample. Albeit there are no physical reasons to believe that these data are unaffected by heating, it appears that for $T_B$=4.2K, the authors claim that `no values of h can be found to fit the experimental R(T) curve satisfactorily even if the specific shoulder structure in the IVC  were not present'. The key to the resolution of this confusion is presented by Fig.1(b) and the corresponding discussion which suggest that the quoted values are seriously underestimated as compared to the effective $T_B$ (see the arrow in Fig.1(b)). The extent of this underestimation could now be evaluated, since the very same $h=350Wcm^{-2}K^{-1}$ should convert these IVC into the very same R(T) hence providing an independent way to estimate the effective $T_B$. Our analysis suggests that these IVCs are actually taken at $T_B\simeq$60 and 65K correspondingly, thus reaffirming quantitatively the values anticipated in Fig.1(b).

Finally, using this $h$=350$Wcm^{-2}K^{-1}$ one can easily quantify the self-heating which builds any IVC point of interest (see Fig.2(b)). The data in this figure suggest that the temperature rise calculated in Ref.\cite{0} under the dubious assumption that the heat dissipated in the sample, sandwiched between metal electrode and substrate of exceptionally poor thermal conductivity, escapes exclusively into the substrate is seriously underestimated.
 
To conclude, it is demonstrated, using exclusively the data from the commented article, that unlike the remarkably consistent state-of-art experiment by \cite{0}, neither the interpretation nor the conclusions are beyond dispute. It is shown that the experimental IVC taken above and below $T_c$ at vastly different $T_B$ spanned over 150K are described quantitatively by Newton's Law of Cooling and Ohm's law using the normal state resistance of the same sample only. This finding confirms the heating origin of the IVC by \cite{0} and suggests that unlike conventional spectroscopy \cite{STM}, the heating in ITS is not a small perturbation but a principal cause of IVC nonlinearity, no matter whether the sample is of centimetre or submicrometre size. 

Our conclusions by no means 
rule out the experimental approach by the authors of \cite{0}. In addition to the remarkable consistency mentioned above, this approach provides a greatly improved heat transfer coefficient which makes it possible reliably to address the intrinsic response and perform other worthwhile IJT experiments, some of which were proposed by \cite{1}. Indeed, the $h=300-350Wcm^{-2}K^{-1}$ estimated above represents a very major improvement, as so far all known experimental $h$ fell into the 2-60$Wcm^{-2}K^{-1}$ range \footnote{It should be noted that the larger mesa by this group reveals a noticeably smaller $h\simeq70Wcm^{-2}K^{-1}$. However, this value is estimated from the single 3-point IVC and hence is of radically lower reliability than the quantities discussed above}. Furthermore, this improvement seems to be robust as the quantitatively similar $h$=300$Wcm^{-2}K^{-1}$ is revealed by another sample (of somewhat smaller $A=0.09\mu m^2$) by this group, \cite{009}. 

\centerline{\bf Acknowledgement}

I am grateful to the authors of Ref.\cite{0} for reaffirming the findings in Fig.1(b) with as-measured data and for admitting the consistency of our explanation hence providing strong independent evidence in support of our basic conclusions.

\end{document}